# Parasitic modes suppression in CW cold tests of 1.3 GHz 9-cell high Q cavities at IHEP

Zhenghui Mi [a,b,c,d], Feisi He [a,b,c], Weimin Pan [a,b,c, d]*, Peng Sha [a,b,c], Jiyuan Zhai [a,b,c], Xuwen Dai [a,c], Song Jin [a,b,c], Zhanjun Zhang [a,c], Chao Dong [a,b,c], Baiqi Liu [a,b,c], Hui Zhao [a,c], Rui Ge [a,b,c], Jianbing Zhao [a,c], Zhihui Mu [a,c], Lei Du [a,b,c], Liangrui Sun [a,b,c], Liang Zhang [a,c], Conglai Yang [a,c], Xiaobing Zheng [a,c], Haiying Lin [a,b,c], Guangwei Wang [a,b,c], Xiangcong He [a,b,c,d]

[a]Institute of High Energy Physics, Chinese Academy of Sciences, Beijing 100049, China
[b]Key Laboratory of Particle Acceleration Physics & Technology, Institute of High Energy Physics, Chinese Academy of Sciences, Beijing 100049, China
[c]Center for Superconducting RF and Cryogenics, Institute of High Energy Physics, Chinese Academy of Sciences, Beijing 100049, China
[d]University of Chinese Academy of Sciences, Beijing 100049, China

**Abstract**

The CW RF test of 1.3 GHz 9-cell cavity in liquid helium bath at 2 K is a very important key point in the cavity procurement. Some problems can be found through the test, according which to optimized and improve the process of cavity. Recently, Medium temperature (mid-T) furnace bake of 1.3 GHz 9-cell cavities have been carried out at IHEP. Through the proceed of mid-T bake, the quality factor of cavity has been greatly improved. While the excitation of the parasitic modes in the high Q cavities CW cold test has been encountered, which implies an error source for the cavity gradient and quality factor determination. In order to ensure the testing accuracy of superconducting cavity, we have improved the testing system. Finally, the parasitic mode is completely suppressed and the CW RF cold test of high Q cavity is guaranteed.

## 1. Introduction

Nowadays, Superconducting radio-frequency (SRF) cavities have been adopted by many accelerators all over the world [1-5]. In order to reduce the construction cost and operational cost of the cryogenics system, Q value of SRF cavities have been increased gradually through many methods in the recent years. The medium temperature (mid-T) furnace bake of 1.3 GHz 9-cell cavities have been carried out at IHEP, through which the Q has been greatly improved. As Fig. 1 shown is vertical test results of EP baseline and mid-T furnace bake of 1.3 GHz 9-cell cavities.

During vertical cavity tests of the 1.3 GHz 9-cell high Q cavities, the spontaneous excitation of parasitic modes with resonance frequencies close to the main mode is frequently observed, especially the 7/9 PI mode. the 7/9 PI mode with a frequency of 1297 MHz is excited and it grows exponentially with a time constant that depends on the quality factor. It has a high-quality factor (Q of order $10^{10}$) due to a weak external coupling. As Fig. 2 shown is an excitation of 7/9 PI mode changes $Q_0$ and Eacc. The calculate value Q increased and the Eacc decreased during the PI mode and 7/9 PI mode together, due to the power meter measures two modes, then the errors are larger [6].

---

∗Corresponding author:
   *Email address*: panwm@ihep.ac.cn (Weimin Pan)

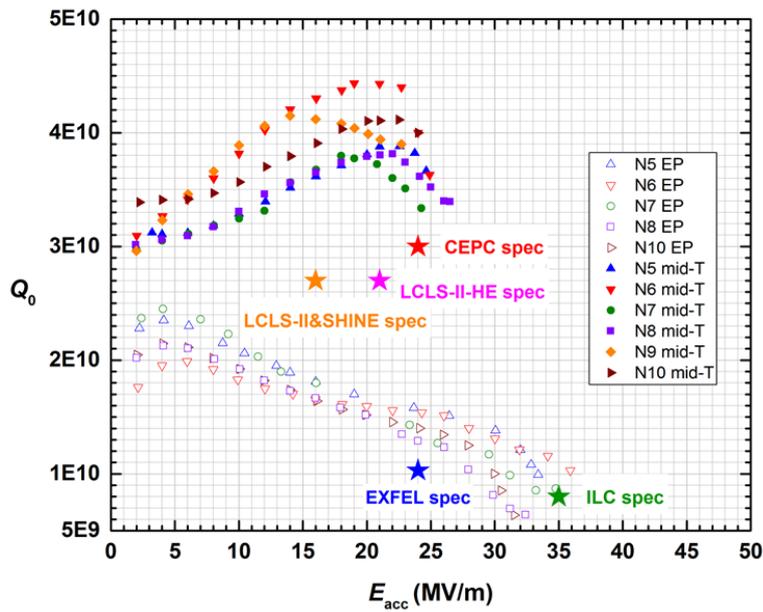

**Fig. 1** Vertical test results of six 1.3 GHz 9-cell cavities (solid icon: mid-T furnace bake; hollow icon: EP baseline)

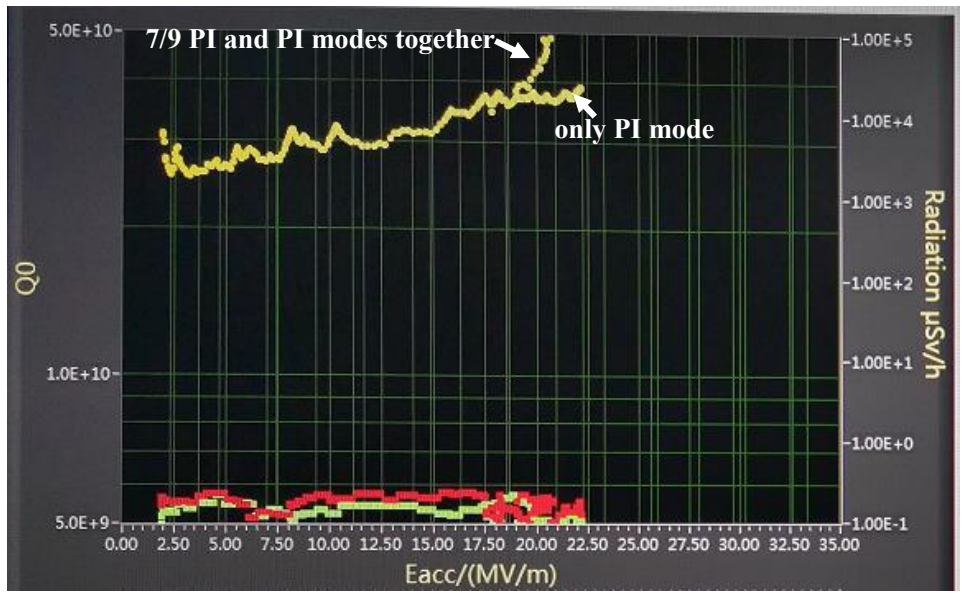

**Fig. 2** An excitation of 7/9 PI mode changes $Q_0$ and $E_{acc}$

The parasitic generation is detected by the Spectrum Analyzer, which through RF cable connected to the pickup antenna, as figure 3 shown.

In order to deal with this problem, most laboratories through turning RF on and off to avoid the interfere of parasitic modes [7]. But the time window for taking reliable data is short and also some problems we cannot found in the shorter test time. We hope that the gradient and Q of the superconducting cavity can be maintained for a long time during the CW RF cold test to check its reliability and stability, and find out the problems existing in it. Therefore, we investigated and analyzed the causes of the parasitic modes excitation, and made targeted improvements to the radio frequency system, successfully suppressed the parasitic mode, and enabled the high Q superconducting cavity to maintain a stable gradient and Q value for a longer time.

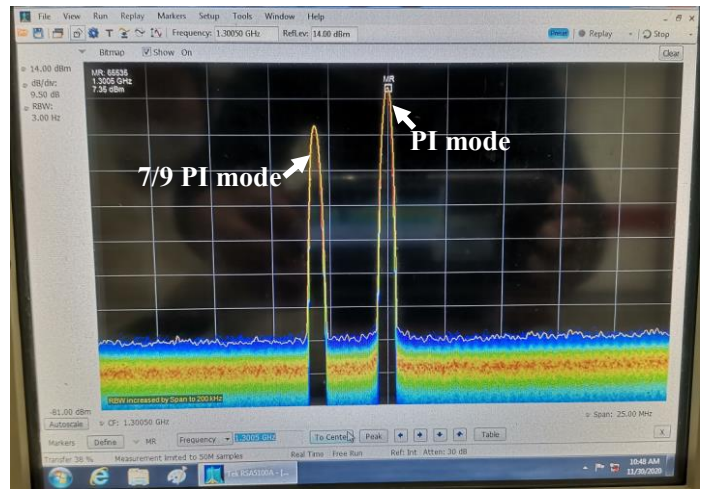

Fig. 3 PI mode and 7/9 PI mode observed by spectrum analyzer

## 2. Causes of parasitic mode excitation

The parasitic modes are generally excited by field emission electrons. The experimental phenomena and results are consistent with HZB's theory. During the testing process of high cavity, the 7/9 PI mode interference begins to appear from about 15M/m. For the 7/9 PI mode, little bremsstrahlung is observed outside the cryostat, due to the radiation energy is low, which is at the order 100 ~ 200 keV. Measured bremsstrahlung on axis (shielded only by the cavity wall) yields energies up to 50 keV. The spontaneous excitation of the 7/9 PI parasitic mode suggests that the electrons are accelerated to a high energy and then are again decelerated to low impact energy, transferring their energy to the 7/9 PI mode. The power fed to the mode is relatively high, on the order of dozens of watts. Such power can only be transferred by the low emission currents of some microamperes if electrons reach the energy of the order of MeV and before being decelerated again. This transfer of energy is feasible due to a modulation of the beam trajectories and the emitted charge by the fields of the 7/9 PI mode [8,9].

## 3. Principle of parasitic modes suppression

Due to the parasitic modes are generally excited by field emission electrons, and the mode excitation depends strongly on the $Q_{load}$. So, there are three methods to dissolve this problem: change the input antenna coupling; repeat HPR to clean the cavity; through the RF direct feedback system to suppression. Repeat HPR the cavity is the fundamental way, but needs more time and the assembly process also maybe cause pollution. In order to test more cavities, fixed input antennas are used, so we can not change the input coupling during cold test, otherwise the test accuracy will also be affected. Therefor we add the RF direct feedback system (DRFB) to the low level rf system, to suppress the parasitic modes. Through the DRFB reduces the equivalent impedance seen by the beam, change the Qload of the beam-cavity system [10].

As figure 4 shown is the test system schematic diagram with parasitic modes suppression function. The test system consists of the following parts: 1-RF power source; 2-Directional coupler; 3-Absorbing load; 4-Power meter; 5-SRF cavity; 6-Bandpass filter; 7-Limiting amplifier; 8-Attenuator; 9-Electronically controlled phase shifter; 10-Coupler; 11-Phase shifter; 12-Frequency meter; 13-Spectrometer; 14-Oscilloscope; 15-Power dividers; 16-Circulator; 17-Power combiner; 18-Input antenna; 19- Pickup antenna; 20-Low pass filter.

In this scheme, the cavity vertical test parasitic modes suppression is realized by adding direct feedback self-excited control loop on the basis of the general measurement method to suppress the other modes of superconducting cavity. In Figure 4, the part within the black dotted line frame is the general self-excited loop. The part outside the black line frame with the power source, directional coupler, circulator and superconducting cavity of the general self-excited loop constitute the direct feedback self-excited loop.

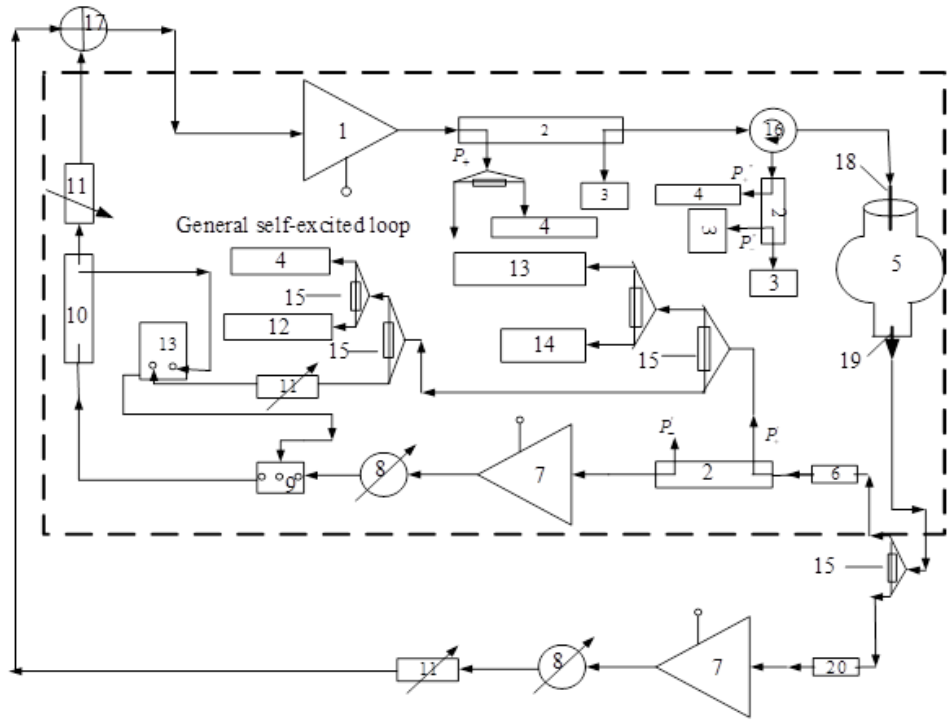

Fig. 4 Test system schematic diagram

As shown in Figure 2, each device is connected by RF coaxial cable. The microwave signal extracted from 1.3 GHz 9-cell cavity by pickup antenna is divided into two channels through power divider. One signal along the general self-excited loop, and the other signal enters the direct feedback self-excited control loop. From the general self-excited loop get the PI mode with parasitic modes, meanwhile only get the parasitic modes through the DRFB loop, the two signals are synthesized to drive the power source, then form a complete self - excited loop test system. The general self-excited loop and the direct feedback self-excited loop in the process of superconducting cavity testing work simultaneously. By adjusting the phase shifter in the direct feedback control self-excited loop, the parasitic modes caused by field emission electrons can be effectively suppressed.

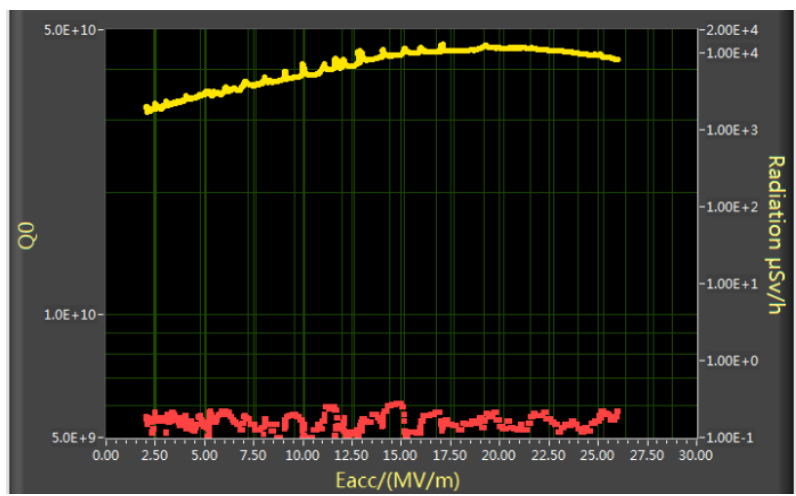

Fig. 5 Cavity test curve with the parasitic modes suppression system

As figure 5 shown is a 1.3 GHz 9-cell high Q cavity test curve with the parasitic modes suppression system at IHEP. During the CW RF cold test, the 7/9 PI parasitic mode would appear at about 18 MV/m without DRFB loop, while the parasitic mode can be suppressed when the DRFB loop open. Also, during the pass band test of 9-cell cavity, other parasitic modes can be successfully suppressed.

4. **Conclusion**

Through add DRFB loop to the general vertical test system, the parasitic modes have been successfully suppressed during 1.3 GHz 9-cell high Q cavity tests. It's helpful to test the performance of cavity stably. But in order to ensure the performance of the superconducting cavity, it is necessary to ensure that the superconducting cavity clean in the process of processing and assembly.

**References**

[1] H.T. Hou, J.F. Chen, Z.Y. Ma et al., Prototypes fabrication of1.3 GHz superconducting RF components for SHINE project, Proceeding of SRF2019, Dresden, Germany, 2019.
[2] J. Y. Zhai, D. J. Gong, H. J. Zheng et al., Design of CEPC superconducting RF system, International Journal of Modern Physics A 34, 1940006 (2019).
[3] J. N. Galayda (ed.), LCLS-II Final Design Report, LCLSII-1.1-DR-0251-R0 (2015).
[4] M. Ball, A. Burov, B. Chase, A. Chakravarty, A. Chen, S.Dixon, J. Edelen, A. Grassellino, D. Johnson, S. Holmes, S.Kazakov, A. Klebaner, I. Kourbanis, A. Leveling, O. Melnychuk,D. Neuffer, T. Nicol, J. Ostiguy, et al., The PIP-II Conceptual Design Report (2017).
[5] D. Reschke, V. Gubarev, J. Schaffran et al., Performance in the vertical test of the 832 nine-cell 1.3 GHz cavities for the European X-ray Free Electron Laser, Phys. Rev. Accel. Beams 20, 042004.
[6] G. Kreps, A. Gossel, et al., EXCITATION OF PARASITIC MODES IN CW COLD TESTS OF 1.3 GHZ TESLA-TYPE CAVITIES, Proceedings of SRF2009, Berlin, Germany.
[7] S. Posen, High Gradient Module Project, International Workshop on Future Linear Colliders, LCWS2021.
[8] V. Volkov, J. Knobloch, A. Matveenko et al., Monopole passband excitation by field emitters in 9-cell TESLA-type cavities, PHYSICAL REVIEW SPECIAL TOPICS - ACCELERATORS AND BEAMS 13,084201(2010).
[9] S.Noguchi. Parasitic Mode Excitation. TTC Meeting Orcay (2009).
[10] C. Zhang, Beam loading effection and RF direct feedback, Second BEPCII Brightness Enhancement Seminar, 2011.